# Remote sensing of cloud base charge

R. Giles Harrison*, Keri A. Nicoll* and Karen L. Aplin[+]
*Department of Meteorology University of Reading, UK
[+]Department of Physics, University of Oxford, UK
e-mail: r.g.harrison@reading.ac.uk

*Abstract*—**Layer clouds are abundant in the Earth's atmosphere. Such clouds do not become sufficiently strongly charged to generate lightning, but they show weak charging along the upper and lower cloud boundaries where there is a conductivity transition. Cloud edge charging has recently been observed using balloon-carried electrometers. Measurement of cloud boundary charging without balloons is shown to be possible here for low altitude (<1km) charged cloud bases, through combining their effect on the surface electric field with laser time of flight cloud base height measurements, and the application of simple electrostatic models.**

## I. Introduction

Thunderstorms are widely appreciated to be strongly electrified clouds, but it is less widely known that other clouds also carry charge, although to a lesser extent. Horizontal layer clouds containing liquid water are abundant across the planet and affect the balance of incoming and outgoing radiation which determines atmospheric temperature. Even without the mixed phases of water and turbulent motions which are key ingredients of thunderstorm electrification, liquid water layer clouds become charged at their upper and lower boundaries. This is due to the vertical current flow always present in the atmosphere through the cloud from the global atmospheric electric circuit, which is also the source of the atmospheric electric field always observed in cloudless fair weather conditions. The importance of droplet charging in such clouds comes from theoretical considerations indicating an enhancement of droplet coalescence, and may ultimately shorten the timescale to rain formation.

Layer cloud edge charging results from the interaction of the vertical current with the vertical conductivity transition at the cloud edge. The current flowing is essentially a global parameter, and the cloud edge properties a local meteorological parameter. Charging of the upper and lower horizontal edges of extensive layer clouds has been observed at sites in both hemispheres using balloon-carried electrometers, demonstrating negative charge in the cloud base and positive charge at the cloud top [1]. Opportunities for such measurements are however limited, and each balloon sounding provides only a single pass through a cloud.

For low altitude clouds, the charge in the cloud base can affect the electric field measured at the surface. If the distance from the surface to the cloud is known, the charge in the cloud base can be determined. Such a remote sensing approach for cloud base charge is evaluated here, using cloud and surface electric field measurements obtained at the University of Reading Atmospheric Observatory.

## II. ATMOSPHERIC MEASUREMENTS

Surface measurements of the vertical atmospheric electric field $E_z$, are, by convention, recorded as the Potential Gradient ($-E_z$) which is positive in fair weather. At Reading University Atmospheric Observatory, the Potential Gradient (PG) is measured continuously using an all-weather field mill (JCI model JCI131), together with cloud base height using an infra-red laser time-of-flight ceilometer (Vaisala model CL31). Geometrical distortions on the electric field measured by the field mill are corrected to that of an open site by comparison with measurements from a stretched wire antenna, which is long-established standard method [2, 3]. The ceilometer retrieves the amount of backscattered power at different time intervals after a pulse of laser light is emitted upwards, allowing a vertical profile of the atmosphere's properties beneath a cloud to be imaged: the greatest backscatter is generated by the cloud droplets, which is used to determine the position of the cloud edge and the cloud base height (see fig 1a).

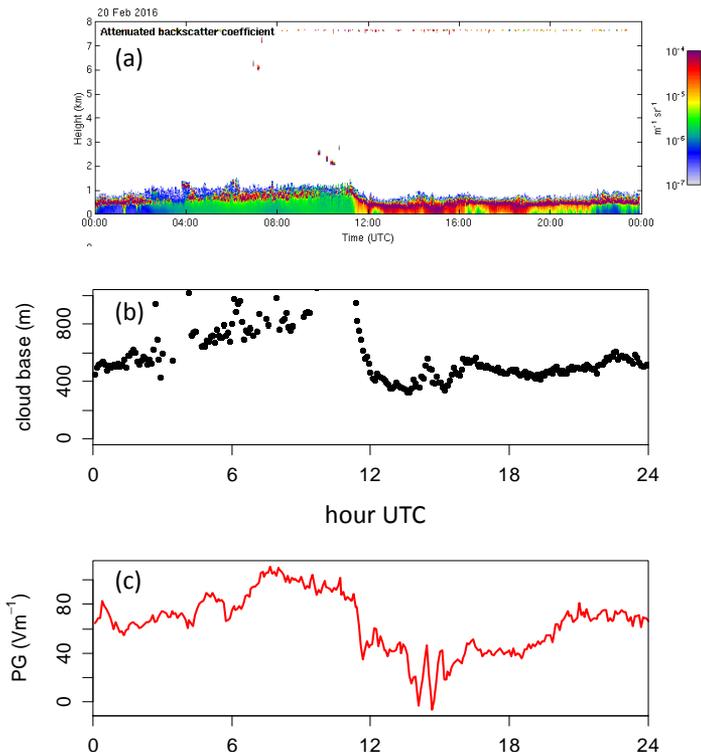

Fig. 1. Time series (plotted against hour of day) from 20[th] February 2016 at Reading, during persistent cloud with cloud base below 1000 m. (a) Backscatter retrieved using a CL31 ceilometer, (b) cloud base height derived by the ceilometer algorithm from the backscatter and (c) Potential Gradient (PG) measured at the surface using a JCI131 field mill. (The cloud base height and PG were recorded as 5 minute average values.)

For extensive and persistent layer clouds having cloud base heights below about 1 km, the PG and cloud base height can be well correlated (fig 1b and c). This because the changes in height occur more slowly than the timescale for the cloud base charge to change, and therefore the electrostatic effect is similar to that of a fixed charge being moved towards and away from the measuring sensor. Changes in cloud base height can result from a variety of causes, but commonly they are associated with variations in the daily cycle of solar heating, with timescales of hours. In comparison, the timescale of the cloud edge charging is relatively rapid, and typically of order 200 to 500s [4].

### III. Electrostatic Considerations

The correlation between fig 1b and fig 1c can be understood in terms of the induced effect of the steady cloud base charge. Consequently, as the cloud base lowers, the surface PG will be suppressed by the negative charge in the cloud base. Conversely, as the cloud base rises, the suppressing effect on the PG is reduced, until the cloud base charge is sufficiently distant for no effect to be observable and the PG returns to its undisturbed fair weather value. This response can be modelled in the simplest case by assuming a single fixed charge of some appropriate geometry in the cloud base. Several possible representations can be considered, such as a point charge, line charge and charged disk (fig 2).

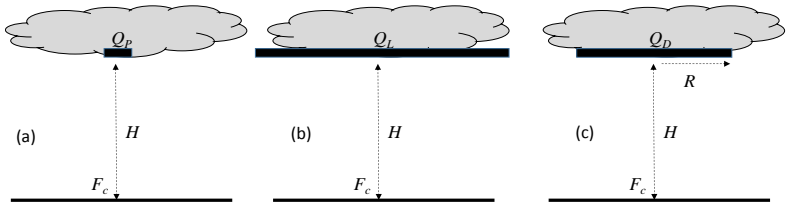

Fig 2. Representation of an extensive layer cloud containing cloud base charge inducing a potential gradient $F_c$ at the surface. The cloud base charge is alternately regarded as (a) a point charge $Q_P$ at height $H$, (b) part of an infinite line charge $Q_L$ at $H$, and (c) a disk charge at $H$ of charge density $Q_D$ with finite radius $R$.

The reduction $F_c$ of the surface PG associated with the cloud base charge at height $H$ is given in the point charge representation by

$$F_c = \frac{Q_P}{4\pi\varepsilon_0 H^2} \quad (1),$$

in the line charge representation by

$$F_c = \frac{Q_L}{2\pi\varepsilon_0 H} \quad (2)$$

and the disk charge representation by

$$F_c = \frac{Q_D}{2\pi\varepsilon_0}\left[1 - \frac{H}{\left(H^2 + R^2\right)^{1/2}}\right] \quad (3).$$

The corresponding point, line and disk charges are given by $Q_P$, $Q_L$ and $Q_D$ respectively, with $R$ the radius of the disk charge. $\varepsilon_0$ is the permittivity of free space, which is con-sidered a good approximation for the medium, as, even in cloudy air, the fraction by mass of water vapour is small.

## IV. RESULTS

Plotting the ceilometer measurements of cloud base height against those of the simultaneously obtained Potential Gradient allows the electrostatic effect to be investigated further, fig 3. This clearly shows the effect on the PG of the cloud base approaching the surface, but as diminishing effect on the PG as the cloud base moves away. At cloud base heights of about 1000m, an influence of the cloud base charge is no longer detectable, and the PG approaches that of a typical fair weather value.

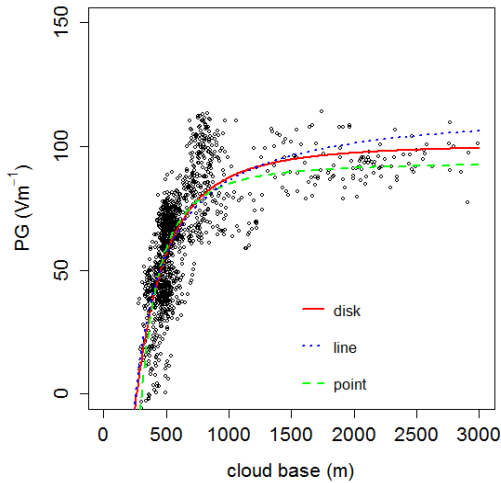

Fig. 3. Cloud base height plotted against simultaneous, co-located measurements of vertical potential gradient (PG) at the surface. (Measurements for 20th February 2016 at Reading determined as 1 min averages, yielding 1349 values with $H$ between 200m and 3000m). Model fits for the point, line and disk charge representation are also included.

Equations (1) to (3) have been fitted to the data in fig 3 using the non-linear regression functions of the R statistical package [6]. In each case the fits were made allowing for the final asymptotic value (i.e. the undisturbed potential gradient $F_0$) as a free parameter. For the point and line charge cases, two parameters were therefore determined, and in the disk charge case, three parameters. Table 1 shows the parameters found for each model, together with the standard errors on the fit: all the fits were statistically significant at $p<0.001$ or better.

TABLE 1: FITTED PARAMETERS

| Model | $F_0$ (Vm$^{-1}$) | Cloud base charge | Cloud disk radius (m) |
|---|---|---|---|
| Point charge | 93.5 ± 0.8 | -(0.96± 0.02) mC | - |
| Line charge | 116.3 ± 1.3 | -(1.7 ± 0.04) µC m$^{-1}$ | - |
| Disk charge | 101.0 ± 1.4 | -(4.8 ± 0.6) nC m$^{-2}$ | 329 ± 32 |

## V. Discussion

The cloud base charge found by the different charge models all provide fair representation of the variations in the data, and confirm the original assumption that the cloud base charge can effectively be regarded as constant on the timescale of the cloud height variations. However, the point charge model is somewhat unphysical for a cloud base where it is known theoretically that the charge is distributed along the lower cloud edge and hence the line or disk charge is more appropriate. Neither of these is entirely satisfactory though, as, in the line charge case, the requirement for an infinite uniform surface is clearly not met, and, in the disk charge case, there is an additional parameter (the disk radius) to be determined, so the fit is more difficult to obtain statistically.

Nevertheless, an advantage of the disk model is that it is bounded, and provides an estimate of the surface charge density in the cloud base. This can then be compared with the previous balloon-borne measurements which provide the volumetric charge density. Further, as the cloud base will often show some structure due to wave structures and turbulent processes, the existence of a finite length scale as required is physically plausible. For wind speeds at the cloud base of ~ 1 ms$^{-1}$ and charging times of ~ 500 s, the associated length scale of ~ 500 m is comparable with that determined for $R$ from the three parameter fit in Table 1. The parameters of the disk charge model can therefore be regarded as physically the most useful in remote sensing of cloud base electrical properties.